\documentclass[RNAAS0]{aastex63_cranmer}
\usepackage{times}

\begin{document}

\title{Updated Measurements of Proton, Electron, and Oxygen
Temperatures in the Fast Solar Wind}

\correspondingauthor{Steven R. Cranmer}
\email{steven.cranmer@colorado.edu}

\author[0000-0002-3699-3134]{Steven R. Cranmer}
\affiliation{Department of Astrophysical and Planetary Sciences,
Laboratory for Atmospheric and Space Physics,
University of Colorado, Boulder, CO 80309, USA}

\begin{abstract}
The high-speed solar wind is typically the simplest and least stochastic
type of large-scale plasma flow in the heliosphere.
For much of the solar cycle, it is connected magnetically to
large polar coronal holes on the Sun's surface.
Because these features are relatively well-known (and less complex than
the multiple source-regions of the slow wind), the fast wind is often a
useful testing-ground for theoretical models of coronal heating.
In order to provide global empirical constraints to these models, 
here we collect together some older and more recent measurements of
the temperatures of protons, electrons, and oxygen ions as a function
of radial distance.
\end{abstract}


\section{}

\begin{figure}[p!]
\begin{center}
\includegraphics[scale=0.95,angle=0]{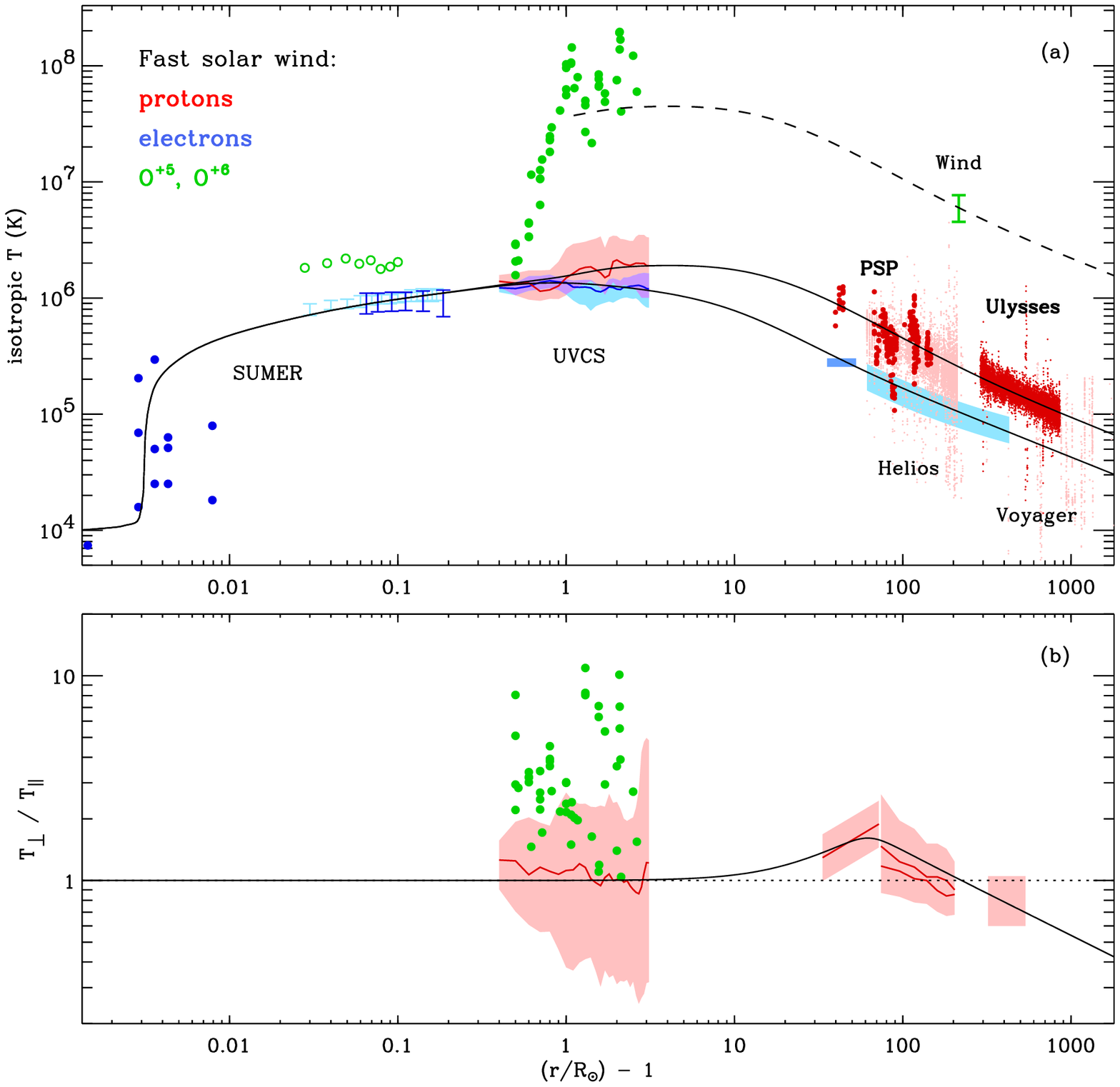}
\caption{(a) Electron data (in blue) from left to right:
SUMER on-disk ``correlation heights'' in open funnels \citep{Maet06},
SUMER off-limb temperatures from
\citeauthor{Wm06} (\citeyear{Wm06}; dark blue) and
\citeauthor{La08} (\citeyear{La08}; light blue),
UVCS \citep{Cr20}, PSP \citep{Ha20}, and Helios/Ulysses \citep{Cm09}.
Proton data (in red) from left to right:
UVCS \citep{Cr20}, PSP (dark red), Helios (light red),
Ulysses (dark red), and Voyager (light red)
high-speed wind data.
Oxygen-ion data (in green) from left to right:
SUMER off-limb kinetic temperatures \citep{LC09},
UVCS \citep{Cr08}, and Wind \citep{Co96}.
(b) From left to right, O$^{+5}$ and proton anisotropy data from
UVCS \citep{Cr08,Cr20},
PSP data with $v \geq 450$ km~s$^{-1}$ \citep{Hu20},
Helios data with $v \geq 600$ km~s$^{-1}$ \citep{Ma82a},
and Ulysses data with $v \geq 600$ km~s$^{-1}$ \citep{Mt07}.
All black curves are solely to guide the eye.
\label{fig01}}
\end{center}
\end{figure}

The plasma in coronal holes and the fast solar wind is not in thermal
equilibrium.
The relatively low density results in a low frequency of particle-particle
collisions, and this allows the electrons and individual ion species to
all exhibit potentially different kinetic properties
\citep[see][]{Ma06,Ve19}.
For example, both in~situ and remote-sensing measurements have found
preferential heating of ions relative to electrons,
systematic differences in bulk flow speeds, and a range of
velocity-distribution anisotropies---i.e., unequal temperatures measured
in the directions parallel and perpendicular to the background magnetic
field.
This research note presents a summary of the radial dependences of
measured temperatures and field-aligned anisotropies in high-speed
wind regions.
Figure \ref{fig01} of this note is an update of
Figure 6 of \citet{Cr09} and
Figure 5(b) of \citet{Cr17} with some new and reanalyzed data.
Specifically, Figure \ref{fig01}(a) shows isotropic temperatures $T$,
defined as $(T_{\parallel} + 2 T_{\perp})/3$, and Figure \ref{fig01}(b)
shows anisotropy ratios $T_{\perp}/T_{\parallel}$.

Data from the Solar Ultraviolet Measurements of Emitted Radiation
(SUMER) instrument on the Solar and Heliospheric Observatory (SOHO)
show trends for the electron temperature $T_e$ near the solar surface.
It is likely that the height of the sharp transition region
fluctuates rapidly in both time and as a function of position.
O~VI 103.2, 103.7~nm line widths from \citet{LC09} show hints of
preferential ion heating at low heights.
At larger distances, data from the Ultraviolet Coronagraph Spectrometer
(UVCS) on SOHO begin to show even stronger departures from thermal
equilibrium (i.e., $T_{\rm ion} \gg T_p \gtrsim T_e$).
These data were obtained over the north and south polar coronal holes
during the 1996--1997 solar minimum.
\citet{Cr20} reanalyzed H~I Ly$\alpha$ profiles with a model that
included mass and momentum conservation, which enabled the thermal
motions of protons and electrons to be separated from one another and
from ``nonthermal'' transverse-wave motions.
The light-colored regions shown in Figure~\ref{fig01}(a) are $\pm 1 \sigma$
bounds on the median values shown with dark-color curves.
\citet{Cr08} analyzed UVCS O~VI profiles, but they did not
separate thermal from nonthermal wave motions.
That was done here using Equation~(24) of \citet{Cr20}, and for
simplicity the error-bars from \citet{Cr08} are omitted.

In~situ data from Parker Solar Probe (PSP), Helios, and Voyager were
obtained from NASA's Space Physics Data Facility (SPDF), and data
from Wind and Ulysses were extracted from previously published sources
\citep[e.g.,][]{Co96,Cm09}.
For most missions, we plot temperature data only for solar-wind speeds
$\geq 600$ km~s$^{-1}$.
However, since PSP has not flown through very much high-speed wind,
the criterion was reduced to 500 km~s$^{-1}$ for $T_p$ and
450 km~s$^{-1}$ for $T_{p \perp}/T_{p \parallel}$ \citep{Hu20}.
One-hour-average data from the PSP Solar Probe Cup \citep{Ka16}
was obtained for the first five perihelia (i.e., times between
2018 October and 2020 August).
Out of 3,742 one-hour intervals with good $u_p$ and $T_p$ data,
the fast streams (with $v \geq 500$ km~s$^{-1}$) comprised only
242 intervals (about 6.5\% of the total).
Reported radial temperatures are interpreted as $T_{p \parallel}$, and
we derive the isotropic $T_p$ using the anisotropy fit shown in
Figure~\ref{fig01}(b).

The black curves in Figure~\ref{fig01} are meant only to guide the eye,
but they are illustrative of coarse trends.
They were originally based on the one-fluid semi-empirical model
used by \citet{CvB12}.
With the modifications made here, $T_e$ peaks at
1.35~MK, at a heliocentric distance of
$r = 1.84 \, R_{\odot}$, and decreases monotonically outwards.
The $T_p$ curve keeps increasing until a peak value
of 1.91~MK at $r = 5.12 \, R_{\odot}$.
At larger distances, the
ratios $T_p / T_e$ and $T_{\rm O} / T_p$ approach
values of about 2.5 and 23.5, respectively.
The latter value indicates more than mass-proportional heating for the
oxygen ions, which was noted by \citet{Co96} for the fast wind.
At distances $\gtrsim 0.2$~AU,
the fits give $T_p \propto r^{-0.70}$ and $T_e \propto r^{-0.61}$.
If one assumes a polytropic index $\gamma$, such that
$T \propto r^{2(1 - \gamma)}$ in a spherically expanding wind, then
this fit implies $\gamma = 1.35$ for protons and 1.31 for electrons.
In Figure~\ref{fig01}(b), one can see that at the largest heights,
$T_{p \perp} / T_{p \parallel} \propto r^{-0.41}$.
After solving for $T_{p \perp}$ and assuming a radial magnetic field
($B \propto r^{-2}$), the magnetic moment $\mu = T_{\perp}/B$ is
roughly proportional to $r^{1.25}$, indicating ongoing
perpendicular heating in interplanetary space \citep[see also][]{Ma06}.

There is much additional work that could be done to supplement these
plots.
PSP will be getting closer to the Sun in the coming years.
Also, kinetic properties of helium (i.e., alpha particles) have
been observed extensively in~situ \citep{Ma82b,Rei01},
and there exists the possibility of extracting these properties from
remote-sensing data as well \citep[e.g.,][]{Mo20}.
Next-generation UVCS-type instruments may be able to measure the
kinetic properties of dozens of different ions and to detect subtle
departures from Gaussian line shapes that point to specific
non-Maxwellian velocity distributions.

\acknowledgments

The author gratefully acknowledges Bernhard Fleck for the
inspiration to compile the most up-to-date data into a new figure.
We acknowledge use of PSP, Helios, and Voyager plasma data acquired
from the NASA/GSFC Space Physics Data Facility's CDAWeb.
PSP was designed, built, and is operated by Johns Hopkins/APL
as part of NASA’s LWS program (NNN06AA01C).
UVCS is a joint program of the Smithsonian Astrophysical Observatory,
Agenzia Spaz\-i\-ale Italiana, and the Swiss contribution to
ESA's PRODEX program.
For a more comprehensive list of UVCS acknowledgments see \citet{Ko95}.

\end{document}